\documentclass[onecolumn,showpacs,amsmath,amssymb]{revtex4}
\input{epsf}

\usepackage{graphicx}
\usepackage{array}
\usepackage{rotating,booktabs}
\usepackage{booktabs,threeparttable}
\begin{document}

\title{B-spline one-center method for molecular Hartree-Fock calculations }

\author{Shi-lin Hu$^{[a,b]}$, Zeng-xiu Zhao$^{[c]}$, and Ting-yun Shi$^{[a]}$} \affiliation{$^{a}$ State Key Laboratory of Magnetic Resonance and Atomic
and Molecular Physics, Wuhan Institute of Physics and Mathematics,
Chinese Academy of Sciences, Wuhan 430071, People's Republic of
China\\$^{b}$ University of Chinese Academy of Sciences, Beijing 100049, People's Republic of China\\$^{c}$ Department of Physics, National University of Defense
Technology, Changsha 410073, People's Republic of China}

\begin{abstract}
We introduce one-center method in spherical coordinates to carry out Hartree-Fock calculations. Both the radial wave function and the angular wave function are expanded by B-splines, and the radial knots and angular knots are adjusted to deal with cusps properly, resulting in the significant improvement of convergence  for several typical closed-shell diatomic molecules. B-splines could represent both the bound state and continuum state wave function properly, and the present approach has been applied to investigating ionization dynamics for H$_2$ in the intense  laser field adopting single-active-electron model.
\end{abstract}

\maketitle



\section*{Introduction} 

There has been a growing interest in the multi-electron dynamics when atoms or molecules interact with intense laser field, due to
the development of laser technology.\cite{Gordon2006,Zhao2007,Boguslavskiy2012,Krausz2009} As is well known, the
computational task grows in exponential way as the degrees of  freedom for the system increase. To overcome this difficulty, theoretical physicists have paid much attention to  the time-dependent Hartree-Fock(TDHF) method \cite{Kulander1987,Nikolopoulos2007,Isborn2008,Lotstedt2012} recently, since it has the promise for simulating electron dynamics of extended systems and takes account of the response of all electrons in the strong laser field. In general TDHF scheme, the time-dependent wave functions are expanded in terms of a complete set of time-independent Hartree-Fock orbitals. To achieve meaningful predication from TDHF, both high-quality bound and continuum Hartree-Fock orbitals are needed. So it's necessary to perform accurate Hartree-Fock calculations for atoms and molecules.

There are a number of approaches to study electronic properties of diatomic molecules within Hartree-Fock approximation for various applications. Roothaan describes molecular orbitals with a linear combination of atomic orbitals(LCAO), and obtains the molecular orbitals and orbital energies by linear variational method.\cite{Roothaan1951}  However, the atomic orbitals are usually expanded by Slater-type or Gaussian-type functions, which suffer from linear dependence, and the basis parameters need to be chosen carefully. After that, many kinds of numerical Hartree-Fock methods arise. Partial-wave self-consistent-field method is developed to decrease the total energies of N$_2$ and CO by more than $10^{-3}$ a.u. than those obtained by LCAO methods.\cite{McCullough1975,Christiansen1977} Meanwhile, finite element method and finite difference method are  applied to diatomic molecule calculations, and good accuracy is achieved.\cite{Heinemann1988,Heinemann1990,Laaksonen1986,Kobus1996,Kobus2012} Compared to the traditional LCAO method,
finite difference method is more accurate because of basis independence. Another approach for accurate Hartree-Fock calculation of diatomic molecules  adopts B-splines and associated Legendre polynomials, and  the second-order correlation effect is calculated by perturbation theory.\cite{Artemyev2004}  Recently, the Hartree-Fock calculations for diatomic molecules have been carried out by Hermite spline associating with collocation method.\cite{Morrison2000,Morrison2009} However, the above-mentioned numerical methods are carried out in spheroidal coordinates, which it is not easy to include nuclear motion or extend to study the electronic properties for multinuclear linear molecules.\cite{Plummer,Haxton2011}

One-center method and B-spline basis have been widely used to investigate diatomic molecular structures.\cite{Bachau2001,Apalategui}
The main advantages of one-center method are that the calculation of diatomic molecule is similar to atomic calculation, and it is natural to take account of nuclear motion and investigate the electronic properties of multinuclear linear molecules.\cite{Goldman1998,Abu-samha} $3\times10^{-5}$ a.u. accuracy has been reached for the lowest state of H$_{2}^{+}$
 by one-center method, in conjunction with the properties of B-spline.\cite{Brosolo1993} The accuracy of $7\times
10^{-8}$ a.u.  has been yielded for the ground state of H$_2^{+}$, which is based on infinite angular expansions and
one-center approach.\cite{Goldman1998}  Resonant photoionization of H$_2$ has been investigated by one-center scheme and B-spline functions,\cite{Sanchez1999} since the continuum states are well represented. Bian et al. have employed one-center method and B-spline basis to study ionization dynamics of the one-electron systems H$_{2}^{+}$ and H$_{3}^{2+}$ in intense laser field.\cite{Bian2008a,Bian2008b} Recently, one-center method has been combined with B-spline functions to study hydrogen molecular ion in strong magnetic field, which could rival multi-center calculations.\cite{Zhang2012}

The purpose of this work is to extend one-center method and B-spline basis to investigate the structure of many-electron atoms and molecules at Hartree-Fock level, and lay the foundation for the development of TDHF.   To our knowledge, this is the first attempt to expand radial wave function and angular wave function by B-spline functions in the molecular Hartree-Fock calculations within the framework of one-center method. The advantages are twofold. First, the evaluation  of two-electron integrals is time-consuming in Hartree-Fock theory, while B-splines  are local basis functions,\cite{Bachau2001} so it could save much time to calculate two-electron integrals. The second one is that radial B-splines  could handle cusps by adopting coincident knots at the nuclear positions,\cite{Zhang2012} and angular B-splines could speed up convergence than spherical harmonic functions.\cite{Kang2006}

 The organization of the present paper is as follows. Firstly, the Hartree-Fock formulas associating one-center method  with B-spline basis and single-active-electron(SAE) model are briefly described  in section II; secondly, the results and discussions of some typical atoms and diatomic molecules are shown in section III; finally, our conclusions and outlooks are presented in section IV.

\section*{Methods}

\subsection*{Hartree-Fock formulas}

Positioning the origin of the spherical coordinates at middle point between the two nuclei, within the Born-Oppenheimer approximation,
the total Hamiltonian of closed-shell  diatomic molecule with 2n electrons could be written as follow(Atomic units are employed
 throughout this article.):

\begin{equation}\tag{1}
H_{2n}=\sum_{i}^{2n}\big[-\frac{1}{2}\nabla^{2}_i-\frac{Z_a}{|\textbf{r}_i-\frac{\textbf{R}}{2}|}-
\frac{Z_b}{|\textbf{r}_i+\frac{\textbf{R}}{2}|}\big]+\frac{1}{2}\sum_{i\neq
j}\frac{1}{r_{ij}}+\frac{Z_aZ_b}{R}.
\end{equation}

\begin{figure}
\includegraphics{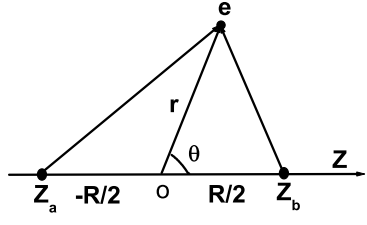}
\caption{\label{fig1} Coordinates of the diatomic molecule.
}
\end{figure}

The first term includes the electron kinetic energies and the coulomb potentials between the electron and the nuclei.  $Z_{a}$ and $Z_{b}$ are the charges of the two nuclei, and R is the internuclear distance. $\frac{1}{r_{ij}}$ is the repulsion potential
between electrons. The third term is constant, which indicates the interaction between the two nuclei. The coordinates of  diatomic molecules are shown in Fig.1. The spatial molecular orbital $\Psi_{i}$ could be expanded by basis functions $\varphi_{\nu}(\textbf{r})$:

\begin{equation}\tag{2}
\Psi_{i}(\textbf{r})=\sum_{\nu}C_{\nu}^{i}\varphi_{\nu}(\textbf{r}).
\end{equation}

The total energy of the system is given by:\cite{Roothaan1951}
\begin{equation}\tag{3}
E=\sum_{i}^{n}\sum_{\nu}\sum_{\mu}C_{\nu}^{i}C_{\mu}^{i}(2H_{\nu
\mu}+2J_{\nu\mu}-K_{\nu \mu}),
\end{equation}
where  $J_{\nu\mu}$ indicates the electron-electron coulomb interaction and $K_{\nu\mu}$ denotes the electron-electron exchange interaction.

\begin{equation}\tag{4}
H_{\nu\mu}=<\varphi_{\nu}|H|\varphi_{\mu}>,
\end{equation}
and
\begin{equation}\tag{5}
H=-\frac{1}{2r^2}{\frac{\partial}{\partial
r}r^2\frac{\partial}{\partial
r}+\frac{1}{2r^2}[\frac{\partial}{\partial
\xi}(1-\xi^2)\frac{\partial}{\partial\xi}-\frac{m^2}{1-\xi^2}]}-Z_a\sum_{\lambda=0}^{\lambda_{max}}\frac{r_<^{\lambda}}{r_>^{\lambda+1}}P_{\lambda}(\xi)
-Z_b\sum_{\lambda=0}^{\lambda_{max}}(-1)^{\lambda}\frac{r_<^{\lambda}}{r_>^{\lambda+1}}P_{\lambda}(\xi),
\end{equation}
where $r_{>}$ ( $r_{<}$ ) is the bigger(smaller) one of (r, $\frac{R}{2}$), and $\xi=\cos\theta$($\theta$ is the angle between
$\textbf{r}$ and $\textbf{R}$). m is the magnetic quantum number, and $P_{\lambda}(\xi)$ are the Legendre polynomials.

\begin{equation}\tag{6}
J_{\nu\mu}=\int\varphi^{*}_{\nu}(\textbf{r}_1)\varphi_{\mu}(\textbf{r}_1)d\textbf{r}_1
\sum_{i=1}^{n}\sum_{\nu_1}\sum_{\mu_1}C_{\nu_1}^{*i}C_{\mu_1}^{i}\int\frac{\varphi^{*}_{\nu_1}(\textbf{r}_2)
\varphi_{\mu_1}(\textbf{r}_2)}{\textbf{r}_{12}}d\textbf{r}_2,
\end{equation}
and
\begin{equation}\tag{7}
K_{\nu\mu}=\sum_{i=1}^{n}\sum_{\mu_1}\sum_{\nu_1}C_{\mu_1}^{i}C_{\nu_1}^{*i}\int\varphi^{*}_{\nu}(\textbf{r}_1)\varphi_{\mu_1}(\textbf{r}_1)d\textbf{r}_1
\int\frac{\varphi_{\nu_1}^{*}(\textbf{r}_2)\varphi_{\mu}(\textbf{r}_2)}{\textbf{r}_{12}}d\textbf{r}_2,
\end{equation}
where two-electron interaction operator could be written in the following formula:
\begin{equation}\tag{8}
\frac{1}{\textbf{r}_{12}}=\frac{1}{r_{>}}\sum_{l=0}^{\infty}\frac{r_{<}^{l}}{r_{>}^{l}}P_{l}(\cos\theta_{12})\\
=\sum_{l=0}^{l_{max}}\sum_{m=-l}^{l}\frac{(l-|m|)!}{(l+|m|)!}\frac{r_{<}^{l}}
{r_{>}^{l+1}}P_{l}^{|m|}(\xi_1)
P_{l}^{|m|}(\xi_2)e^{im(\phi_1-\phi_2)}.
\end{equation}

 $r_{>}$($r_{<}$) corresponds to max(min)($r_{1}$,$r_{2}$), and $P_{l}^{|m|}(\xi)$ are associated Legendre polynomials.

Introducing Lagrange multipliers $\epsilon_{i}$, we could obtain the following equation:
\begin{equation}\tag{9}
E_1=\sum_{i}^{n}\sum_{\nu}\sum_{\mu}C_{\nu}^{i}C_{\mu}^{i}(2H_{\nu\mu}+2J_{\nu
\mu}-K_{\nu\mu}-\epsilon_iS_{\nu\mu}),
\end{equation}
and
\begin{equation}\tag{10}
S_{\nu\mu}=\int\varphi^{*}_{\nu}(\textbf{r}_1)\varphi_{\mu}(\textbf{r}_1)d\textbf{r}_1,
\end{equation}
where S$_{\nu\mu}$ is the overlap integral. Variation of E$_1$ with regard to the coefficient $C_{\nu}^i $, we get the following
equation for the $i$th molecular orbital:
\begin{equation}\tag{11}
(\textbf{H}+2\textbf{J}-\textbf{K})\textbf{C}=\epsilon_i\text{S}\textbf{C}.
\end{equation}
where $\epsilon_i$ indicates energy of the $i$th molecular orbital. After obtaining the orbital energies and the molecule orbitals, we could calculate total energy using the following formula:\cite{Roothaan1960}
\begin{equation}\tag{12}
E=\sum_{i}(H_{i}+\epsilon_{i})+\frac{Z_aZ_b}{R},
\end{equation}
where $H_{i}$ is one-electron integral,
\begin{equation}\tag{13}
H_{i}=\sum_{\nu\mu}C_{\nu}^{i}C_{\mu}^{i}H_{\nu\mu}.
\end{equation}

\subsection*{Single-active-electron model}
The time-independent Schr\"{o}dinger equations for the outmost shell electron within Born-Oppenheimer approximation  are
given by:
\begin{equation}\tag{14}
[H+2\widehat{J}(\textbf{r})-\widehat{K}(\textbf{r})]\Psi_{i}(\textbf{r})=E_{i}\Psi_{i}(\textbf{r}).
\end{equation}

\begin{equation}\tag{15}
\widehat{J}(\textbf{r$_1$})\Psi_{i}(\textbf{r$_1$})=\sum_{j\neq
i}^{n}\int\frac{|\Psi_{j}(\textbf{r$_2$})|^{2}}{\textbf{r$_{12}$}}d\textbf{r$_2$}\Psi_{i}(\textbf{r$_1$})
+\frac{1}{2}\int\frac{|\Psi_{i}(\textbf{r$_2$})|^{2}}{\textbf{r$_{12}$}}d\textbf{r$_2$}\Psi_{i}(\textbf{r$_1$}),
\end{equation}
and
\begin{equation}\tag{16}
\widehat{K}(\textbf{r$_1$})\Psi_{i}(\textbf{r$_1$})=\sum_{\mu\neq
i}^{n}\int\frac{\Psi_{j}^{*}(\textbf{r$_2$})\Psi_{i}(\textbf{r$_2$})}{\textbf{r$_{12}$}}d\textbf{r$_2$}\Psi_{j}(\textbf{r$_1$}),
\end{equation}
where Eq.(15) describes the electron-electron coulomb interaction, and Eq.(16) denotes the electron-electron exchange interaction.
$\Psi_{j}(\textbf{r$_2$})$ is the molecular orbital which we obtain  from Hartree-Fock calculations for diatomic molecule.
As above, we solve Eq.(14) and obtain a series of eigenvalues, in which negative eigenvalues and positive eigenvalues correspond to bound states and discretized continuum states, respectively.

The laser field is linear polarized along the molecular axis, and the time-dependent electric field is defined via the vector potential A(t) as E(t)=-$\frac{\partial}{\partial t}$A(t). The time-dependent wave function could be expressed with the
field-free eigenstates as follow:
\begin{equation}\tag{17}
\Phi(\textbf{r},t)=\sum_{j}C_{j}(t)\Psi_{j}(\textbf{r}).
\end{equation}

The time-dependent Schr\"{o}dinger equations for the highest occupied molecular orbital(HOMO) of diatomic molecule in
intense laser field are written as:
\begin{equation}\tag{18}
i\frac{\partial}{\partial t}\Phi(\textbf{r},t)=[H+2\widehat{J}(\textbf{r})-\widehat{K}(\textbf{r})-r\xi E(t)]\Phi(\textbf{r},t).
\end{equation}

 Substituting Eq.(17) into Eq.(18) and projecting onto $\Psi^{*}_{i}(\textbf{r})$, we adopt Crank-Nicolsen method to propagate the time-dependent wave function.\cite{Bachau2001} After we obtain the time-dependent wave function, the ionization probability at the end of the laser pulse is defined as:

\begin{equation}\tag{19}
P_{ion}=1-\sum_{i=1}^{N(E_{i}<0)}|<\Psi_{i}(\textbf{r})|\Phi(\textbf{r},t_f)>|^2,
\end{equation}
where $t_f$ denotes the time at the end of the laser pulse.

\subsection*{B-spline basis}
Considering the axial symmetry of diatomic molecules, both the radial functions and the angular functions are expressed with B-splines
in the following way:
\begin{equation}\tag{20}
\varphi_{\nu}(\textbf{r})=\frac{1}{\sqrt{2\pi}}\frac{B_{\alpha}^{k_r}(r)}{r}B_{\beta}^{k_\xi}(\xi)(1-\xi^2)^\frac{|m|}{2}e^{im\phi}.
\end{equation}

 $B_{\alpha}^{k_r}$(r) and $B_{\beta}^{k_\xi}(\xi)$ are the B-splines of order k$_r$ and k$_\xi$, respectively. There are many advantages for B-spline.\cite{Brosolo1993,Brosolo1992,Brosolo1994,Shi2004,Bachau2001} On one hand, finite B-splines could construct a nearly complete set, and do not suffer from linear dependence. On the other hand, B-splines could provide a very proper representation of bound state wave function and continuum state wave function,\cite{Vanne2004,Zhao2009,Petretti} and the density of continuum states could be adjusted by changing the truncated radius, which has been applied to studying dynamic properties of molecules such as  photoelectron energy spectra and enhanced ionization.\cite{Nikolopoulos2007,Bian2008a,Bian2008b}
The factor $(1-\xi^2)^{\frac{|m|}{2}}$ describes the asymptotic behavior of wave function for $\xi$ close to $\pm 1$, and also eliminates the singularities in Eq.(5) for $m\neq 0$. For radial B-splines, k$_r$-2 multiple knots are positioned at nuclear locations to handle the cusps properly in radial direction, which has been tested in Ref.[32]. The other knots are distributed in exponential way, so we could obtain satisfactory results for the bound states. For radial B-splines, the knots are distributed in the following way:
\begin{equation}\tag{21}
0=r_{1}=\cdot\cdot\cdot=r_{k_{r}-1}<r_{k_r}<\cdot\cdot\cdot<r_{j+1}=\cdot\cdot\cdot=r_{j+k_r-2}=\frac{R}{2}
<\cdot\cdot\cdot<r_{N_1-k_r+3}=\cdot\cdot\cdot=r_{N_1}=R_{max},
\end{equation}
where $N_{1}$ is the total radial knot number and R$_{max}$ is the truncated radius. Except for the knots locating at the nuclear position and on the boundaries, the other knots are distributed as follows:
\begin{equation}\tag{22}
r_{j}=R_{max}\times\frac{e^{\alpha_1x_j}-1}{e^{\alpha_1x_{N_1}}-1},x_j=j\times\frac{R_{max}}{N_1-2k_r-3}.
\end{equation}

The parameter $\alpha_1$ could be adjusted. The angular wave function is expanded by B-splines, since B-splines include more
high angular momentum effects and  accelerate angular convergence dramatically.\cite{Kang2006} For angular B-splines, the knots sequence are written as follows:
\begin{equation}\tag{23}
-1=\xi_{1}=\cdot\cdot\cdot=\xi_{k_{\xi}}<\cdot\cdot\cdot<\xi_{j-1}<\xi_{j}<\xi_{j+1}<\cdot\cdot\cdot<\xi_{n_{\xi}-k_{\xi}+1}
=\cdot\cdot\cdot=\xi_{n_{\xi}}=1,
\end{equation}
where  $n_{\xi}$ is the total number of angular knots. For atoms and homonuclear molecules, the knots in (-1,0) are distributed in the
following way:
\begin{equation}\tag{24}
\xi_{j}=-\sin[\frac{\pi}{2}\cdot(\frac{j-k_{\xi}-1}{N_2+1})^{\alpha_2}],
\end{equation}
where  $N_2$ is the number of knots in (-1,0). The parameter $\alpha_2$ is adjusted to let more knots position close to
$\pm 1$ to cope with cusps in angular direction, and the knots in [0,1] are in symmetric distribution comparing with the knots in [-1,0].

For heteronuclear molecules, the knots in (-1,0) are distributed as follow:
\begin{equation}\tag{25}
\xi_{j}=-\sin[\frac{\pi}{2}\cdot(\frac{j-k_{\xi}-1}{N_{3}+1})^{\alpha_3}],
\end{equation}
and the knots in (0,1) are distributed in the following sequence:
\begin{equation}\tag{26}
\xi_{j}=\sin[\frac{\pi}{2}\cdot(\frac{j-k_{\xi}-1}{N_{4}+1})^{\alpha_4}],
\end{equation}
where $N_{3}$ is the number of knots in (-1,0) and $N_{4}$ is the number of knots in (0,1). Both the parameters  $\alpha_{3}$ and
$\alpha_{4}$  are adjusted to assure more knots to be distributed around $\pm 1$. In our work, the nucleus with more charge is positioned at -1 in the angular direction, so more knots are distributed in (-1,0) comparing with the range (0,1). However, the knots are distributed as Gauss-Legender-quadrature-points in Ref.[30,31], so the B-spline knots can not be flexibly adjusted.

\section*{Results and Discussions}

In this section, we present the numerical results of some atoms, homonuclear molecules(H$_{2}$ and N$_{2}$) and
heteronuclear molecules(LiH and CO). We choose H$_{2}$ and LiH, because the electronic properties  have been studied in many
different ways and the results could be severed as benchmarks to check our approaches.\cite{Laaksonen1986,Heinemann1988} We select N$_{2}$ and CO, since they have been widely studied in experiments to investigate ionization,\cite{Litvinyuk2003,Lee2011} and we will extend the present approach to study the ionization dynamics for N$_{2}$ and CO in intense laser fields.

 For diatomic molecules, the orders of both the radial B-spline and the angular B-spline are k$_{r}$($k_{\xi}$)=9; for atoms, the order of the radial B-spline is k$_{r}$=7 and the order of the angular B-spline
is k$_{\xi}$=3. $\lambda_{max}$ and $l_{max}$ are 300 and 30, respectively. All the integrals are calculated with Gauss-Legendre
integration. The truncated radius R$_{max}$ is equal to 14 a.u. and 20 a.u. for diatomic molecules and atoms(He and Ne), respectively(unless noted). For Be atom, R$_{max}$ is 40 a.u., since the outmost orbital is not tightly bound.
 The orbitals of one-electron Hamiltonian H are chosen as the initial wave functions, and the
 orbital energies  are used as convergence criterion for the Hartree-Fock iterations. In each iteration, we use the orbitals which
 are obtained the previous time as the trial wave functions.
\subsection*{He, Be and Ne atoms}
Recently, Hartree-Fock codes for atoms have been developed, in which the radial wave function and angular wave function are expanded by B-spline and spherical harmonic functions, respectively.\cite{Saito,Fischer} To test the validity of the present method, we
first calculate the orbital energies and the total energies of He, Be and Ne atom, and show the converged results in Table I.
In Eq.(24) $\alpha_2=0.9$. In all of the Tables, N$_{r}\times$ N$_\xi$ means that N$_r$ radial B-splines and N$_\xi$ angular B-splines are used, and we employ x2dhf program in Ref.[16] to calculate the reference data. Comparing to the accurate Hartree-Fock results,\cite{Kobus2012} we obtain high accuracy for the total energies using small basis number, and the results are quite stable with increasing basis numbers. For He, Be and Ne atom, the
total energies are at least converged to 11 significant digits, and the orbital energies are in good agreement with those in Ref.[16],
which are obtained by finite difference method in prolate-spheroidal coordinates.

\begin{table}
\begin{tabular}{ccccccccccccccccccc}\hline\hline
& \textbf{N$_r\times$N$_{\xi}$} & \textbf{1$\sigma$(1s)} & \textbf{2$\sigma$(2s)} &
\textbf{3$\sigma$(2p)} & \textbf{1$\pi$(2p)} & \textbf{E}\\\hline
& &  & &He& & &  \\
 & 15$\times$4 &  -0.91795558148   &  &     &    & -2.86167995855 & \\
 & 15$\times$6& -0.91795558148  & & & & -2.86167995855 &\\
 &20$\times$4& -0.91795556285 & & & & -2.86167999552&\\
 &25$\times$8&-0.91795556284 & & & & -2.86167999560 & \\
&[16]& -0.91795556287 &  & & & -2.86167999562 &\\\hline
 & &  & &Be& & &  \\
 & 25$\times$4 & -4.7326698972    &  -0.3092695515  &     &    & -14.5730231673 & \\
 & 25$\times$6 &  -4.7326698972  & -0.3092695515& & &  -14.5730231673 &\\
 &30$\times$4&  -4.7326698972 & -0.3092695515& & & -14.5730231680&\\
 &35$\times$8&  -4.7326698972 & -0.3092695515 & & & -14.5730231681 & \\
 &[16]& -4.7326698974 & -0.3092695515 & & & -14.5730231683 &\\\hline
 & & & &Ne& & &  \\
 & 25$\times$4 & -32.772442780    & -1.930390875  & -0.850409652 & -0.850409652&  -128.547098057& \\
 & 25$\times$6 & -32.772442780 & -1.930390875& -0.850409652&-0.850409652 &-128.547098057 &\\
 &30$\times$4& -32.772442791 & -1.930390878& -0.850409650& -0.850409650&  -128.547098106&\\
 &35$\times$8& -32.772442792 &-1.930390879 &-0.850409650 &  -0.850409650&-128.547098108 & \\
 &[16]& -32.772442793 &-1.930390879 & -0.850409650&-0.850409650 & -128.547098109
 &\\\hline\hline
\end{tabular}
\caption{\label{tab:1}Convergence of the orbital energies and the
total energies of He, Be and Ne atom with increasing basis
numbers.}
\end{table}

\subsection*{Homonuclear molecules: H$_{2}$ and N$_{2}$}
In Table II, we give the orbital energies and the total energies of H$_2$  with increasing basis number. The distance between the two
nuclei is 1.4 a.u., and $\alpha_2$ in Eq.(24) is 1. Comparing the results obtained by $60\times 38$ and $65\times 42$ basis numbers, the total energy is converged to 7 significant digits. Comparing the total energy for basis numbers 65$\times$42 with that in Ref.[16], the relative difference is about $5.8\times 10^{-7}$.  In Ref.[43], Hartree-Fock calculation is
 implemented with LCAO approach. The good accuracy of total energy is achieved because the wave functions at cusp positions are well described, and the 1$\sigma_{g}$ wave function along the Z axis has been plotted in Fig.2. We find that there are two cusps at the nuclear locations, and the wave function is symmetric with the internuclear midpoint.

\begin{table}
\begin{tabular}{cccccccccccc}\hline\hline
&\textbf{N$_r\times$N$_{\xi}$} & \textbf{1$\sigma_g$} & \textbf{E} & \\\hline
 &$50\times 30$ & -0.59465779 & -1.13362748 &  \\
&$50\times 38$ &-0.59465819 & -1.13362865  & \\
 &$60\times 38$ & -0.59465820& -1.13362866 &\\
 &$65\times 42$ & -0.59465828  & -1.13362892  & \\
 &[16]  & -0.5946585691 & -1.133629571469  &\\
 &[43]  & -0.5946585687  &  -1.133629571456 & \\\hline\hline
\end{tabular}
\caption{\label{tab:2}Convergence of the orbital energies and the
total energies for  H$_2$  as a function of basis
number.}
\end{table}

\begin{figure}
\includegraphics{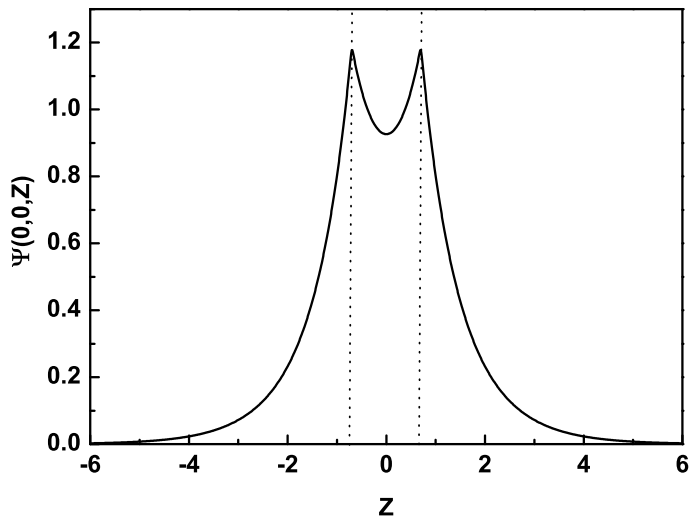}
\caption{\label{fig2} The 1$\sigma_{g}$ wave function along  the
Z axis for H$_{2}$ with internuclear distance R=1.4 a.u.
}
\end{figure}

In Table III, we display the convergence test of orbital energies and the total energy of N$_2$ with increasing basis-set sizes. The
equilibrium distance  R=2.068 a.u. is assumed, and $\alpha_2$ in Eq.(24) is 0.4. Comparing the results obtained by $50\times 38$ and $55\times 42$ basis sizes, the total energy converges to 5 significant digits. For basis-set sizes 55$\times$42, the relative error of the total energy is about $ 2.8\times 10^{-5}$ compared with that in Ref.[16] and the
orbital energies agree well with those from Ref.[16]. For $55\times 42$ basis sets, each iteration takes about forty minutes in the Intel i3-2120 CPU with 1 GB of RAM, and it needs 32 iterations to obtain the above results.

\begin{table}
\begin{tabular}{ccccccccccccc}\hline\hline
 &\textbf{N$_{r}\times$N$_{\xi}$}& \textbf{1$\sigma_g$} & \textbf{1$\sigma_u$} & \textbf{2$\sigma_g$} & \textbf{2$\sigma_u$} & \textbf{3$\sigma_g$} & \textbf{1$\pi_u$} & \textbf{E} &\\\hline
 & 40$\times$30 &  -15.67943 & -15.67581 &-1.47329 &-0.77797&-0.63484
 &-0.61565&-108.9796 &\\
 &40$\times$38  &-15.68044&-15.67683  &-1.47323  &-0.77802 & -0.63482&-0.61556&-108.9837&
 \\
 & 50$\times$38 &-15.68142  &-15.67780 & -1.47327  &-0.77806
 &-0.63481&-0.61555 &-108.9884 &\\
 & 55$\times$42 &-15.68170 &-15.67808 &-1.47340 &-0.77806 &-0.63480& -0.61562&-108.9908&\\
 &[16] &-15.681866 &-15.678251  &-1.473422  &-0.778077 &-0.634793&-0.615625&-108.99383&  \\
 \hline\hline
\end{tabular}
\caption{\label{tab:3}Same as Table 2, but for  N$_2$.}
\end{table}

\subsection*{Heteronuclear molecules: LiH and CO}
 In Table IV, we present the orbital energies and the total energies of LiH  with increasing basis numbers. $\alpha_3$ in Eq.(25) and $\alpha_4$ in Eq.(26) are 0.42 and 0.75, respectively. Comparing the results obtained by $60\times
38$ and $65\times 42$ basis numbers, the total energy converges to 5 significant digits. For basis numbers 65$\times$42, the
relative difference of the total energy is around $4.6\times 10^{-6}$ in comparison with that in Ref.[16]. The internuclear distance R=3.015 a.u., which is much bigger than that of H$_2$. The charge distribution is not symmetrical either. So it is difficult to obtain good accuracy of total energy comparing with H$_{2}$. The total energy 7.987351 a.u. (R=3.0141 a.u.) has been obtained by Artemyev et.al, which is better than our result, since B-splines are combined with prolate spheroidal coordinates to handle cusps naturally for diatomic molecules.\cite{Artemyev2004}

\begin{table}
\begin{tabular}{cccccccccccccccc}\hline\hline
 & \textbf{N$_r\times$N$_{\xi}$} & \textbf{1$\sigma$}  & \textbf{2$\sigma$} & \textbf{E} &\\\hline
 & $50\times 30$ &  -2.445188  & -0.301725  & -7.987198 & \\
 & $50\times 38$ &  -2.445212  & -0.301724 &  -7.987259 &  \\
& $60\times 38$ &  -2.445221 & -0.301725 & -7.987282 &\\
& $65\times 42$ &-2.445216 & -0.301734& -7.987315&\\
&[16]&-2.44523386 & -0.30173837&-7.98735238 & \\\hline\hline
\end{tabular}
\caption{ \label{tab:4}Convergence of the orbital energies and the
total energies for  LiH with increasing basis sizes.}
\end{table}

In Table V, we present the convergence test of orbital energies and the total energy of CO with increasing basis numbers. The
internuclear distance R=2.132 a.u. $\alpha_3$ in Eq.(25) and $\alpha_4$ in Eq.(26) are 0.43 and 0.5, respectively. Comparing the results obtained by $50\times 38$ and $55\times 42$ basis sizes, the total energy is at least converged to 5 significant digits. For basis numbers 55$\times$42, the relative difference of the total energy equals to around
$4.2\times 10^{-5}$ comparing with that in Ref.[16]. It is well known that orbital distribution is asymmetrical around nuclei for heteronuclear molecules. To better describe the nonsymmetric distribution of charge distribution for CO, the possibilities of charge
distribution for the $i$th molecular orbital around the nucleus A and nucleus B are defined as:
\begin{equation}\tag{27}
P_{A}=\int_{-1}^{0}d\xi\int_{0}^{2\pi}d\phi\int_{0}^{R_{max}}|\Psi_{i}(r,\xi,\phi)|^{2}r^{2}dr,
\end{equation}
and
\begin{equation}\tag{28}
P_{B}=\int_{0}^{1}d\xi\int_{0}^{2\pi}d\phi\int_{0}^{R_{max}}|\Psi_{i}(r,\xi,\phi)|^{2}r^{2}dr,
\end{equation}
 respectively.
 The ratios of the possibilities for the charge distribution around O nucleus to that
near C nucleus are 3/1, 77/23, 3/1 and 3/22 for the $3\sigma$, $4\sigma$, $1\pi$ and  $5\sigma$ orbitals, respectively.

\begin{table}
\begin{tabular}{ccccccccccccccccc}\hline\hline
 & \textbf{N$_r\times$N$_{\xi}$} & \textbf{1$\sigma$}  & \textbf{2$\sigma$} & \textbf{3$\sigma$} & \textbf{4$\sigma$}
 &\textbf{1$\pi$} & \textbf{5$\sigma$}& \textbf{E} &\\\hline
 & $40\times30$ &  -20.66101 & -11.35609 & -1.52141 &
 -0.80449&-0.64048&-0.55487& -112.7697&\\
 & $40\times38$ & -20.66222 &-11.35837 &-1.52141  &-0.80451
 & -0.64040 &-0.55490 &-112.7776 & \\
 & $50\times38$ &-20.66480 &-11.35869 &-1.52152  &-0.80453
 & -0.64038&-0.55489 &-112.7844 &\\
 & $55\times42$ & -20.66546&-11.35938 &-1.52136 &-0.80448  &-0.64022  &-0.55494&-112.7862 & \\
 &[16] & -20.664521 &-11.360051   & -1.521489  & -0.804529  &-0.640361&  -0.554923&-112.79091 & \\
 \hline\hline
\end{tabular}
\caption{\label{tab:5} Same as Table 4, but for CO.}
\end{table}

\subsection*{\textbf{Intensity-dependent ionization: H$_2$}}
 To further extend the present approach to study complex molecules in the intense laser field, we first employ SAE model to investigate the ionization yields of H$_2$ as a function of laser intensity. There have been some works adopting SAE approach to
 study H$_2$ in the laser field.\cite{Awasthi2005,Awasthi} In Ref.[44], only configurations which one electron dwells in the 1$\sigma_g$ orbital of H$_2^+$ are taken account of. In Ref.[45], the core electron is fixed in the initial orbital obtained by Hartree-Fock method or density-functional theory in the field-free case, and the other electron is influenced by the combined field of the core and the laser pulse. The present SAE model is similar to that in Ref.[45]. In contrast to the present work, the angular wave functions are expanded by spherical harmonics in Ref.[45].

   R$_{max}$ is 120 a.u., and 150 radial B-splines and 16 angular B-splines are used. The initial state is 1$\sigma_g$ state, and the orbital energy is -0.5944 a.u. The vector potential is $A(t)=E_{0}/\omega\cos^{2}(\pi t/t_{max})\cos(\omega t)$,  $-t_{max}/2<t<t_{max}/2$. t$_{max}$ is the duration of the laser pulse with 36, 24 and 12 optical cycles for 266 nm, 400 nm and 800
nm central wavelength, respectively.  During the time propagation of the wave function, a $\cos^{\frac{1}{8}}$ absorber wave function is adopted in range [80,120] to smoothly bring down the wave function near boundary. The time step is $\Delta$t=0.015 a.u., 0.025 a.u. and 0.05 a.u. for 266 nm, 400 nm and 800 nm wavelength, respectively.  Fig.3 shows the ionization probabilities of H$_2$ as a function of laser intensity for various wavelength. The ionization yields become bigger with the increase of laser intensity, and a reasonable agreement is found between our data and those based on Hartree-Fock method in Ref.[45].
\begin{figure}
\includegraphics[width=1.0\textwidth]{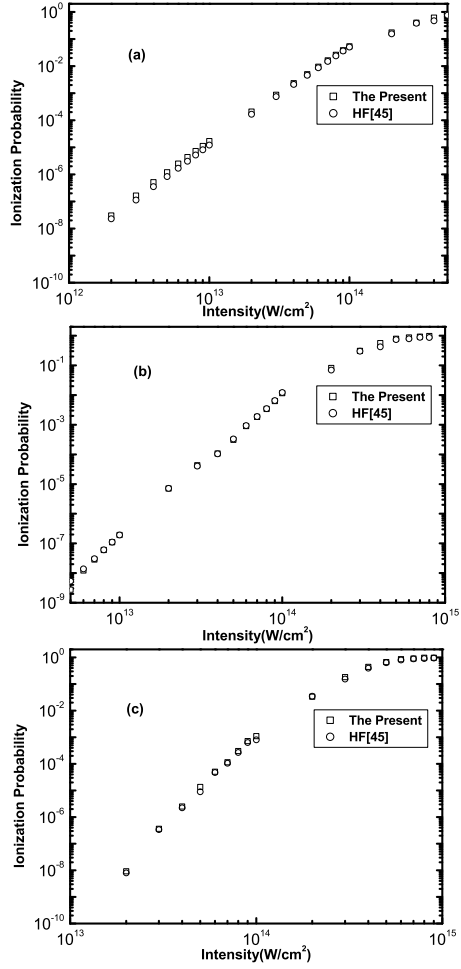}
\caption{\label{stabiliz2}   The ionization probability as a function of the laser intensity for different wavelength, (a)266 nm, (b) 400 nm, (c) 800 nm. }\label{Figure_3}
\end{figure}

\section*{Conclusions and Outlooks}

We have performed calculations of  the orbital energies and total energies for some typical atoms and diatomic molecules at
Hartree-Fock level with one-center method in spherical coordinates. Both the radial wave functions and the angular wave functions are
expressed with B-spline to speed up convergence. The present approach is easy to implement and could obtain good
accuracy  for atomic and molecular calculation. For atoms, the orbital energies and total energies could achieve good convergence
with small basis numbers. For H$_2$ and N$_2$,  the total energies are converged to 7 and 5 significant digits respectively, and the electron cloud distributes evenly around the nucleus; for LiH and CO,  the total energies are converged to 5 and 5 significant digits respectively, and the charge distribution is uneven near the nucleus. SAE model is used to study  ionization yields of H$_2$ in the laser field, and the results have a reasonable agreement with those in Ref.[45]. Moreover, both the bound states and continuum states could be approximated  properly associating one-center method with B-spline basis sets, so the present approach is a good starting point to develop TDHF. Finally, the present method may be extended to study the electronic structures of three atomic linear molecule(CO$_2$) and four atomic linear molecule(C$_2$H$_2$).  We also notice that Serra et.al will adopt finite-size-scaling scheme to investigate critical parameters of large systems based on Hartree-Fock theory, in which the orbitals are expanded by B-spline functions.\cite{Serra2012}

\subsection*{Acknowledgments}

We acknowledge Dr. A. Saenz and Dr. M. Awasthi for providing us data, and thank Dr. J. Kobus for enthusiastic help in the x2dhf program. S. L. H. and T. Y. S are supported by  the National Basic Research Program of China under Grant No. 2010CB832803. Z. X. Z. is supported by the Major Research plan of National NSF of China under Grant No. 91121017.
\clearpage




\clearpage





\clearpage


\clearpage

\end{document}